# On the Performance of Short Binary BCH Codes for Ultra-Low Latency Wireless Communications

Robert Morelos-Zaragoza and Jeffrey Ma, San Jose State University

**Abstract**

In recent years, polar codes have been considered for communication systems that require high reliability and ultra-low latency, such as sixth generation (6G) wireless communications. This paper presents simulation results showing that short binary extended BCH (eBCH) codes with low-complexity decoding outperform polar codes for lengths 64 and 128. In the simulations, polar mapping under additive white Gaussian noise (AWGN) is assumed and ordered-statistics decoding (OSD) of eBCH codes is compared with CRC-aided successive-cancellation list decoding (SCLD-CRC) of polar codes of the same lengths and rates. The results indicate that short-length binary eBCH codes achieve lower average bit error rate values (higher reliability) and thus should be considered as strong candidates in communication systems requiring extremely low latency, i.e., short code lengths of up to 128 bits. The eBCH simulation results are obtained with OSD and reprocessing order equal to one so that complexity is comparable to SCLD-CRC. Specifically, error performances are quantified of length-64 and selected length-128 eBCH codes with order-1 OSD and polar codes with SCLD-CRC for the same rates and lengths. These results serve to verify that short binary eBCH codes do indeed outperform short polar codes with comparable decoding complexity.

# 1. Introduction

This paper considers short binary eBCH codes in situations where very low latency and high reliability are required. The motivation of this work comes from ever-growing low latency and high reliability requirements of wireless communication networks [1]: "Message transfer latency no longer than 100 ms with 20 ms maximum allowed latency in some specific use cases, without any compromise on reliability." It is well known that these requirements can be satisfied using short binary error-correcting codes (ECC). In the past, research in this area had been focused on polar codes (see [2] and references therein) using successive-cancellation list-decoding (SCLD) and SCLD with cyclic redundancy check (SCLD-CRC). However, other types of channel coding schemes appeared to be overlooked. An exception is reference [2], where codes obtained by interpolating between polar codes and Reed-Muller (RM) codes are studied and trade-offs examined between decoding complexity and error performance. Previous papers related to the performance of short eBCH codes include reference [3], which considers several coding schemes, such as low density parity check (LDPC), Polar, tail-biting convolutional code (TB-CC) and Turbo codes, compared with length 128 eBCH codes, and reference [4], where an eBCH (128,64) code is compared with LDPC and turbo codes of the same length and rate.

The main contribution of this paper is the presentation of simulation results to compare the performance of binary ECC of very short length, in particular polar codes and binary eBCH codes, under additive white Gaussian noise (AWGN). We choose extended BCH codes so that they can be compared with polar codes of the same code lengths. The simulation results presented here serve to verify that short length eBCH codes have better performance than short polar codes of the same length and rate and should be considered as strong candidates in applications where low latency is required. Simulations of eBCH codes were obtained with ordered-statistics decoding (OSD) [5] with reprocessing order 1. In the following, $N$ denotes the code length and $K$ the code dimension (or number of information bits). Although decoding complexity depends on many factors, we chose reprocessing order equal to one in OSD so that its complexity, $O(K)$, is relatively close to that of SCLD-CRC, $O(N\log_2 N)$.

## 2. An illustrative example: The performance of a polar code versus an RM code with polar mapping under AWGN with length 64

A construction algorithm (algorithm TR1) of polar codes under AWGN is presented in [6]. Using this algorithm with a 5 dB threshold, it was found that the first instance where RM codes and polar codes differ is at $N=64$ and $K=22$. Notice that this is different from polar code constructions *under erasures*, where the first case in which polar codes and RM codes differ is $N=32$ and $K=16$ [7]. This suggests that polar codes are not the best option for code rates $K/N$ lower than 1/2.

Generator matrices of both codes are depicted in Figure 1 (a dot represents bit value equal to one). Observe that a codeword of Hamming weight (i.e., number of ones) 8 in the generator matrix of the polar code is replaced by a codeword of Hamming weight 16 in the generator matrix of the RM code. This is indicated with a red rectangle in the figure. Since the RM code has a higher minimum Hamming distance, it achieves better performance as shown in the simulation results in Figure 2, which plots the bit error rate (BER) versus the average bit energy to noise ratio ($E_b/N_0$) expressed in decibels (dB).

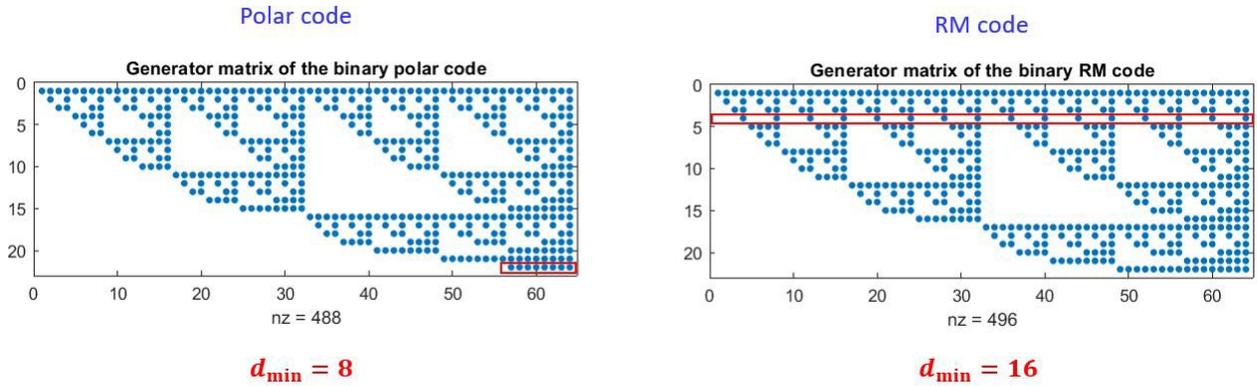

**Figure 1.** Generator matrices of a polar code and an RM code with *N*=64 and *K*=22.

The simulations reported in Figure 2 utilized the following decoding algorithms: (1) Successive cancellation decoding (SCD) [8], successive-cancellation list decoding (SCLD) [9], and (2) SCLD with 6-bit CRC (SCLD-CRC6) ]9], which was found to give best performance among CRC variations, and (3) reprocessing order 5 OSD (OSD5) [10]. The RM code was simulated using OSD5. As expected, this illustrates the well-known fact that polar codes do not perform well at relatively short lengths.

The idea behind this paper is to extend this example to binary BCH codes which are known to outperform RM codes. Binary extended BCH codes, or eBCH codes, are chosen due to their higher minimum distance for the same rate compared to RM codes and for having the same length as polar codes.

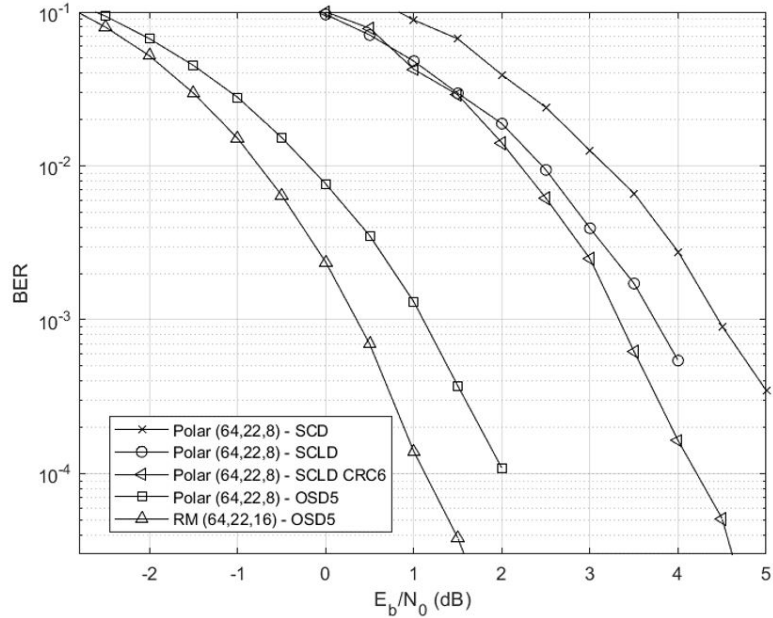

**Figure 2.** Error performance of a polar code and an RM code with $N$=64 and $K$=22.

## 3. Results

*3.1. Binary eBCH and polar codes of length 64*

The performance of OSD with reprocessing order 1 for all binary eBCH codes of length 64 was simulated and the results are shown in Figure 3. SCLD with a 6-bit CRC of type 6-ITU (denoted SCLD-CRC6) was simulated using the AFF3CT software [8] for polar codes with the same length 64 and dimensions. The results are presented in Figure 4. After trying different CRC schemes, a 6-bit CRC was found to give the best performance. The reason for this is that a larger number of CRC bits reduces the coding rate while SCLD yields a smaller reduction in error rate, resulting in an increase in the average bit error rate.

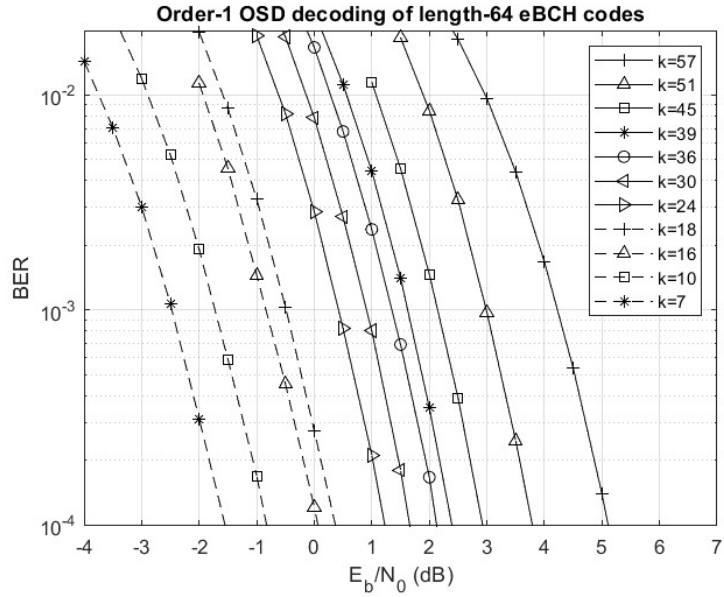

**Figure 3.** Performance of length-64 eBCH codes with order-1 OSD.

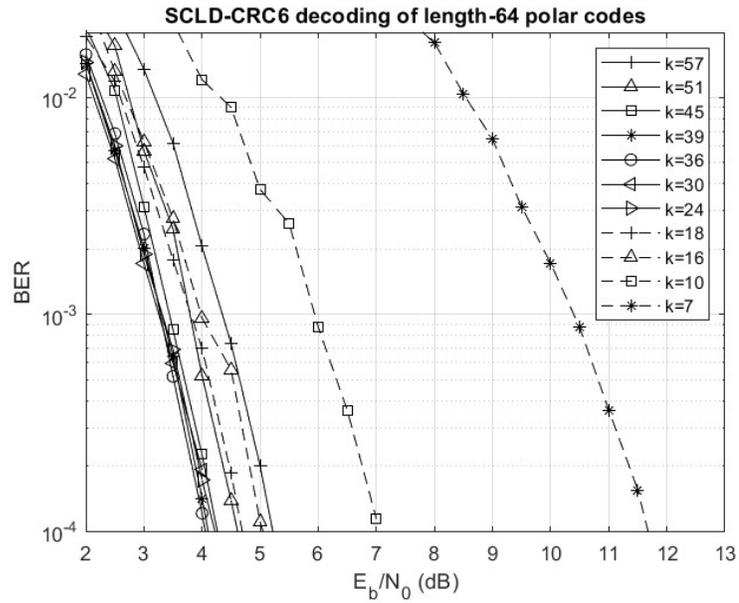

**Figure 4.** Performance of length-64 polar codes with CRC6-aided SCLD.

Table 1 shows the difference Δ in $E_b/N_0$ (dB) required to achieve an average bit error rate of $10^{-3}$. We observe that polar codes of length 64 and rates less than 1/2 (corresponding to dimension less than 32) give poor performance with at least a 2 dB loss compared with eBCH codes. This additional average power is also shown graphically in Figure 5.

**Table 1.** Additional power Δ in dB required by polar codes of length 64 to achieve a $10^{-3}$ BER

| Dimension | Δ (dB) |
|---|---|
| 57 | 0.1 |
| 51 | 0.8 |
| 45 | 1.25 |
| 39 | 1.6 |
| 36 | 1.9 |
| 30 | 2.4 |
| 24 | 2.9 |
| 18 | 4.35 |
| 16 | 4.8 |
| 10 | 7.6 |
| 7 | 12 |

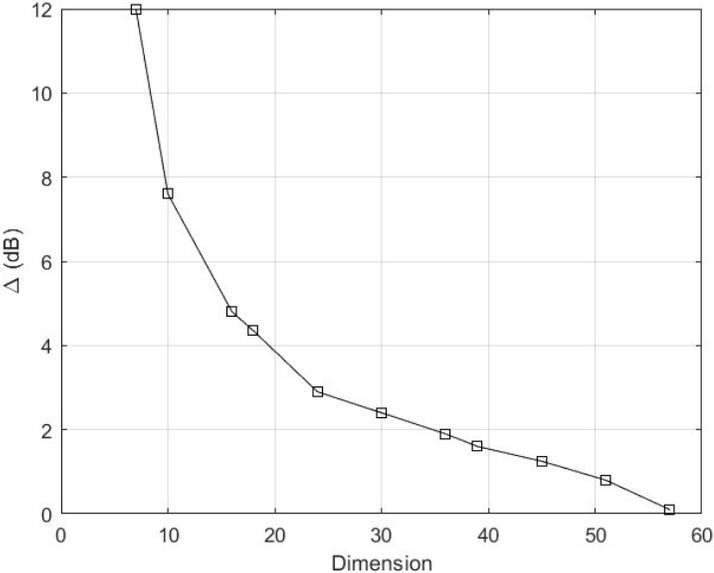

**Figure 5.** Additional power in dB required by length-64 polar codes with SCLD-CRC6.

*3.2. Binary eBCH and polar codes of length 128*

Also simulated was the performance of OSD with reprocessing order 1 for selected binary eBCH codes of length 128. The simulation results are shown in Figure 6. Here SCLD with a 16-bit CRC of type 16-CCITT (SCLD-CRC16) was simulated with AFF3CT [8] for polar codes of the same length and dimensions as the selected eBCH codes. Again, we found this 16-bit CRC scheme to give the best performance among different alternatives. In figure 7 simulation results for polar codes are shown. Once again, the difference $\Delta$ in $E_b/N_0$ (dB) required to achieve an average bit error rate of $10^{-3}$ was computed and is plotted in Figure 8.

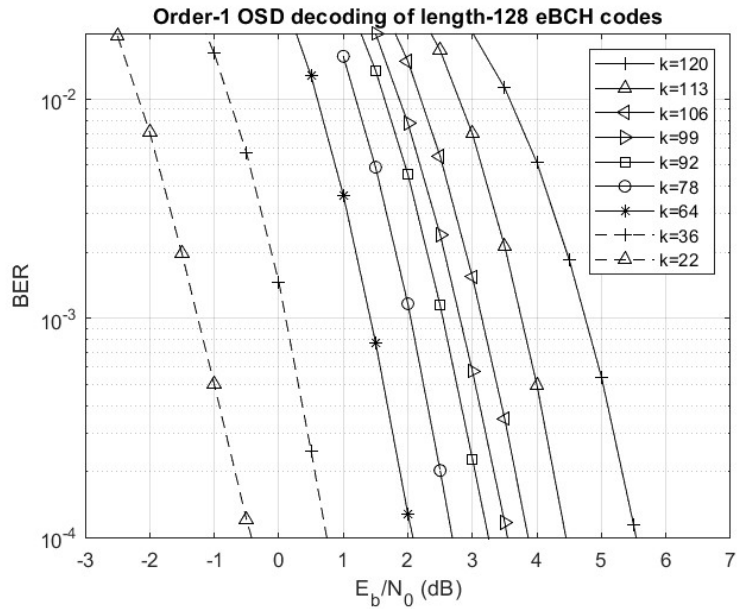

**Figure 6.** Performance of selected length-128 eBCH codes with order-1 OSD.

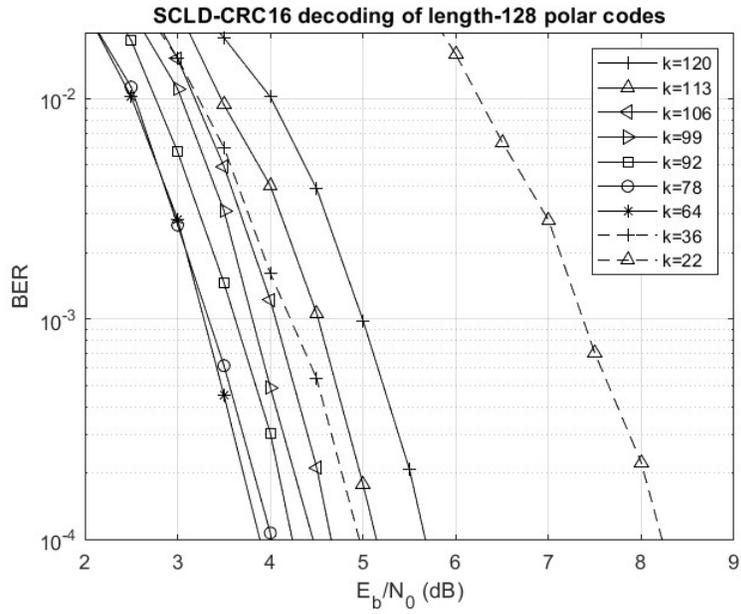

**Figure 7.** Performance of selected length-128 polar codes with CRC16-aided SCLD.

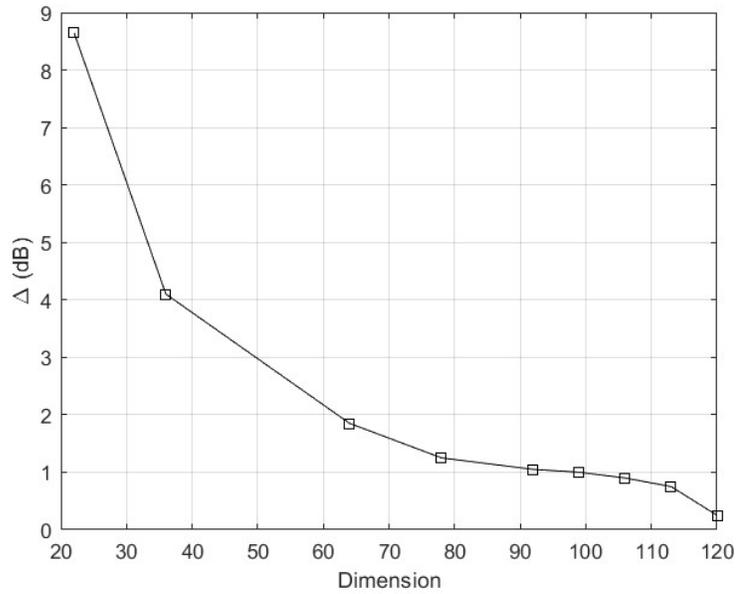

**Figure 8.** Additional power in dB required by length-128 polar codes with SCLD-CRC16.

In the case of $N=128$, polar codes with rates greater than 1/2 exhibit a relatively small difference of up to 1.4 dB in additional average power required to achieve the same error performance. Just as in the case of *N*=64, there is a large loss in average power requirements with polar codes compared to eBCH codes at low coding rates.

## 4. Discussion

In this paper comprehensive simulation results are reported on the performances of length 64 eBCH codes and length 64 polar codes of the same rates. Selected codes of length 128 were also simulated. To reduce complexity, ordered statistics decoding with a reprocessing order equal to 1 was used in the case of eBCH codes, while successive-cancellation list decoding with 6-bit CRC (for *N*=64) and 16-bit CRC (for *N*=128) were used for polar codes. The results indicate that polar codes achieve an error performance similar to eBCH codes at high coding rates, with comparable decoding complexity. It was observed that the performance of polar codes degrades significantly at coding rates lower than 1/2. In all cases studied, eBCH codes outperform polar codes.

While the results presented here for eBCH codes were obtained with reprocessing order-1 OSD, it should be remarked that recently there has been a renewed interest in devising efficient decoding techniques of linear block codes based on bit reliability values, such as Guessing Random Additive Noise Decoding (GRAND) [11] and ordered-reliability direct error pattern testing (ORDEPT) [12]. Moreover, at the time of this writing, the authors became aware of a new algorithm for OSD with low complexity that uses reduced Gaussian elimination that is introduced in [13]. The reported results in this paper serve to further motivate the study of decoding algorithms that trade-off a good error performance (high reliability) and low decoding complexity for short binary eBCH codes and other linear block codes. It should be noted that the exact assessment of complexity depends on many factors, most importantly the implementation platform, see e.g.

[14]. Most recently, the performance of short codes and their decoding algorithms designed specifically for ultra-low latency communication applications is studied in references [15-17].

In summary, a performance comparison has been reported for binary codes of lengths 64 and 128 to achieve ultra-low latency. To the best of the authors' knowledge, this is the first time that the performances of all binary eBCH codes of length 64 have been compared with those of binary polar codes of the same length and rate. It has been shown that short eBCH codes with low-complexity decoding based on ordered statistics outperform short polar codes with SCLD-CRC of the same rates. The difference in performance was found to be relatively small only at very high coding rates, while at low coding rates polar codes with SCLD-CRC suffer high performance degradation. Future research includes the implementation of promising reliability-based decoding algorithms for eBCH codes and more generally for linear block codes, and the evaluation of their decoding complexity with specific hardware and software platforms. Another interesting topic to study further is the amount of CRC bits that are required in SCLD-CRC such that a good balance between reliability and average power requirements (due to a decreased coding rate) is achieved, see e.g. [18] and related references.